\documentstyle[12pt]{article}

\begin{document}
\topmargin 0pt
\oddsidemargin 5mm

\setcounter{page}{1}
\hspace{8cm}{Preprint YerPhI-1480(17),Yerevan,1996}
\vspace{2cm}
\begin{center}

{NONLINEAR TWO-DIMENSIONAL POTENTIAL  PLASMA WAKE WAVES}

{\large A.Ts. Amatuni}\\
\vspace{1cm}
{\em Yerevan Physics Institute}\\
{Alikhanian Brother's St. 2, Yerevan 375036, Republic of Armenia}
\end{center}

\vspace {5mm}
\centerline{{\bf{Abstract}}}

The condition for potential description of the wake waves,
generated by flat or cylindrical driving electron  bunch in cold 
plasma is  derived.

The two-dimensional nonlinear equation for potential valid for small
values of that is obtained and solved by the separation of variables.
Solutions in the form of cnoidal waves,existing behind the moving  bunch
at small values of vertical coordinate,are obtained.In particular,at some
boundary conditions,corresponding to blow-out regime in the underdense 
plasma,the solution represents by a solitary nonlinear wave.

Approximate solution is also obtained using the method of multiple scales.

The indications are obtained that the dependense of the amplitudes on 
longitudinal coordinate determines essentially,even in the first 
approximation,by driving bunch charge distribution.
The wake wave amplitude can increase at some conditions along the 
longitudinal distanse from the rear part of the bunch.

 \section{\small{INTRODUCTION}}
\indent

Analysis of the one dimensional longitudinal,transverse and coupled transverse
longitudinal plain nonlinear waves in cold relativistic plasma are given in the
review \cite{A} (see therein the references on original works).One dimensional 
nonlinear longitudinal waves,generated by the driving bunches with the infinite
transverse dimensions,were considered in \cite{B}-\cite{G}.

In the present work the two-dimensional nonlinear wake waves,generated by the
flat or cylindrical electron bunch,are discussed.

The corresponding linear problem was considered in \cite{H},\cite{I} and was
found,in correspondance with previous result \cite{J},that the magnetic field
in wake wave in linear approximation is equal zero.This result connected with
the absence of the energy flow in the wake wave and the absence of the 
vortexes in plasma electron motion in linear approximation.

In the one dimensional nonlinear treatment \cite{B}-\cite{G} the magnetic field
in the wake wave is also zero (by construction),due to the symmetry of the
problem relative to transverse displasments.

In two-dimensional wake wave the magnetic field is zero only when vortexes
connected with the plasma electron motion are zero,which in this case is an
additional requirement on the type of the motion of the plasma electrons.The
wake waves in this case are potential,i.e. electric field components $E_z,E_y$
can be expressed as a gradient of one scalar function $\varphi(y,z)$,and 
components of the plasma electrons current also can be expressed through one 
scalar function $\psi(y,z)$

The approximate nonlinear equation for 
potential $\varphi$ can be obtained,using Maxwell equations and approximate 
equations  of the motion. Equation for potential has an exact solutions with 
the separated variables for small values of the transverse coordinate.

Among the solutions,which are finite,nonlinear waves by cnoidal nature,
at some boundary conditions,associated with the blow-out regime,
there exists the solution (on separatirix) in form of the solitary wave.

For arbitrary values of the transverse coordinate the approximate 
solution for potential ,using multiple scales perturbative method is found.
 
\section{\small{VORTEX-FREE WAKE WAVE}}
\indent

Consider the wake wave generated in the cold neutral plasma,with the immobile
ions,by the flat electron bunch, which has horisontal dimensions $2a$ much 
larger than vertical dimension $2b$,longitudinal dimension is $2d$.The charge
density in the bunch is $n_b$,electron plasma density is $n_0$,and we consider
both overdense and underdense regimes.

Bunch is moving along $z$-axis with the constant velocity $v_0<c$
in lab system.All the 
physical quantities in the question are considered as a function of vertical
coordinate $y$ and $\tilde{z}=z-v_0t$.An electrical field,generated by the bunch
$|E_x| \ll E_y \neq 0, E_z \neq 0$ and magnetic field $B_z=0, |B_y| \ll 
|B_x| \equiv |B| \neq 0$.

Introduce the dimensionless variables and arguments by
\begin{eqnarray}
\label{AA}
\vec{E}=\sqrt{4{\pi} nmv_{0}^{2}}\vec{E'}=\frac{\omega mv_0}{e}\vec{E'}
\\ \nonumber
\vec{B}=\sqrt{4{\pi} nmv_{0}^{2}}\vec{B'}=\frac{\omega mv_0}{e}\vec{B'}
\end{eqnarray}

\begin{equation}
\label{AB}
z',y' = k\tilde{z},ky, k^2=\frac{\omega^2}{v_{0}^{2}}=
\frac{4{\pi}n{e}^2}{mv_{0}^{2}},n'_b=\frac{n_b}{n},n'_0=\frac{n_0}{n}
\end{equation}
where $n$ is the arbitrary electron density,which is convinient to choose equal
$n=n_b$ in the underdense $(n_b>n_0)$ case and $n=n_0$ in the overdense
$(n_b<n_0)$ case.

Following \cite{K},\cite{H} introduce BFTCh-transformation of the variables
\begin{equation}
\label{AC}
V_z = \frac{{\beta}_{ez}}{\beta-{\beta}_{ez}}, V_y = 
\frac{{\beta}_{ey}}{\beta- {\beta}_{ez}}, 
\beta=\frac{v_0}{c},{\vec{\beta}}_{ez}=\frac{{\vec{v}}_{ez}}{c} 
\end{equation}
\begin{equation}
\label{AD}
n'_e=\frac{{\beta}N}{\beta-\beta_{ez}}=N(1+V_z);
\end{equation}
($N \rightarrow n'_0,\beta_{ez} \rightarrow 0$,
when $z,y \rightarrow +{\infty};v_{ex}=0$).

The Maxwell equations then can be rewritten in the following form 
(superscript prime is ommited in what follows):
\begin{equation}
\label{AE}
\begin{array}{l}

(a) \qquad \displaystyle \frac{\partial B}{\partial y}={\beta}NV_z+
\beta\frac{\partial E_z}{\partial z}+\beta n_b \\
\\
(b)\qquad \displaystyle \frac{\partial (B+{\beta}E_y)}{\partial z}=-{\beta}NV_y 
\\
\\
(c)\qquad \displaystyle \frac{\partial ({\beta}B+E_y)}{\partial z}=
\frac{\partial E_z}{\partial y} \\
\\
(d)\qquad \displaystyle \frac{\partial E_z}{\partial z}+ \frac{\partial E_y}
{\partial y}=(n_0-n_b)-N(1+V_z)
\end{array}
\end{equation}

The continuity equation
$\frac{\partial N}{\partial z}=\frac{\partial NV_y}{\partial y}$
follows from (\ref{AE}.d),(\ref{AE}.a),(\ref{AE}.b).Using (\ref{AE}.a),
(\ref{AE}.b),(\ref{AE}.c) we have
\begin{equation}
\label{AH}
\frac{\partial^2 B}{\partial y^2}+(1-{\beta}^2)\frac{\partial^2 B}
{\partial z^2}=
{rot}_{x}({\beta}N\vec{V})+{rot}_{x}(\vec{\beta}n_b)
\end{equation}
which means that the magnetic field is zero in plasma (linear or nonlinear)
wake wave only when 
\begin{equation}
\label{AI}
rot({\beta}N\vec{V})=0
\end{equation}
i.e. the plasma electrons motion is vortex-free.

In the following,we consider the region of the space,occupied by wake wave
i.e. $z < -d$ .Maxwell equations
(\ref{AE}) for wake waves under condition (\ref{AI}) can be obtained putting in 
(\ref{AE}) $B=0$ and $n_b=0$.
\begin{equation}
\label{AG}
\begin{array}{l}
(a) \qquad \displaystyle \frac{\partial E_z}{\partial z}=-NV_z \\
\\
(b) \qquad \displaystyle \frac{\partial E_y}{\partial z}=-NV_y \\
\\
(c) \qquad \displaystyle \frac{\partial E_y}{\partial z}=\frac{\partial E_z}
{\partial y} \\
\\
(d) \qquad \displaystyle \frac{\partial E_y}{\partial y}=n_0-N 
\end{array}
\end{equation}

Then from (\ref{AG}.c) follows that
\begin{equation}
\label{AJ}
\vec{E}=-grad\varphi
\end{equation}

i.e. the wake fields under condition (\ref{AI}),as it must be,are potential.

\section{\small {THE BASIC EQUATION FOR THE POTENTIAL.EXACT SOLUTION
IN SEPARABLE ARGUMENTS}}
\indent

Consider Maxwell equations (\ref{AG}) for wake waves,when $z < -d$.
From (\ref{AG}.a),(\ref{AG}.d) and (\ref{AJ}) we have
\begin{equation}
\label{AK}
N=n_0+\frac{\partial^2\varphi}{\partial y^2}
\end{equation}

\begin{equation}
\label{AL}
\frac{\partial^2 \varphi}{\partial z^2}=\left(n_0+\frac{\partial^2
\varphi}{\partial y^2}\right)V_z
\end{equation}

From hydrodynamic equation of the plasma wake wave electrons motion,using 
(\ref{AA}),(\ref{AB}),(\ref{AC})
it is possible to obtain the relativistic equation of motion for the $V_z$
component of the generalized velocity:
\begin{eqnarray}
\label{AM}
\frac{-\partial V_z}{\partial z}+V_y\frac{\partial V_z}{\partial y}=-W^{1/2}
\left[E_z\left(1+2V_z+\frac{V_{z}^{2}}{{\gamma}^2}\right)+
{\beta}^{2}V_zV_yE_y\right], \\ \nonumber
W \equiv 1+2V_z+\frac{V_{z}^{2}}{{\gamma}^2}-{\beta}V_{y}^{2};
\end{eqnarray}

Neglecting terms with the squares of generalizid velocity,compared to the terms
with the first power of that,the expression (\ref{AM}) converted to
\begin{equation}
\label{AN}
\frac{\partial V_z}{\partial z} \approx E_z(1+3V_z)
\end{equation}

The solution of this equation,using (\ref{AJ}),is
\begin{equation}
\label{AO}
V_z=\frac{1}{3}(e^{-3\varphi}-1)\approx-\varphi
\end{equation}

with the condition $\varphi=0$,when $V_z=0$.

Substituting (\ref{AO}) in (\ref{AL}) we have the basic equation for $\varphi$
\begin{equation}
\label{AP}
\frac{\partial^2 \varphi}{\partial z^2}+\varphi\frac{\partial^2 \varphi}
{\partial{y^2}}+
n_0\varphi=0
\end{equation}

Nonlinear term in eq. (\ref{AP}) is proportional to $\frac{\partial^2\varphi}
{\partial y^2}$ and can be large.In the linear approximation solution of eq.
(\ref{AP}) lost the $y$-dependence and describes the harmonic oscilation 
with the plasma frequency on $z$;$y$-dependence of the solution comes from
boundary condition at $z=-d$ and coincides with it for all $z < -d$.This 
is always the case,when wake waves are described as the product of the two
functions from separate arguments $y$ and $z$.Such a situation takes 
place in linear approximation \cite{C},\cite{J},\cite{N}.

The eq. (\ref{AP}) permits to search the solution with the separable 
variables
\begin{equation}
\label{AQ}
\varphi(y,z)={\varphi}_1(y){\varphi}_2(z)
\end{equation}
\begin{equation}
\label{AR}
\frac{{\varphi}_2''+n_0{\varphi}_2}{\varphi_{2}^{2}}=
-{\varphi}_1''\equiv -k
\end{equation}

where $k$ is a separation constant.The equations for $\varphi_1$ and 
$\varphi_2$ are:
\begin{equation}
\label{AS}
{{\varphi}_1}''=k
\end{equation}
\begin{equation}
\label{AT}
{{\varphi}_2}''+n_0{\varphi}_2+k{\varphi_2}^2=0
\end{equation}

Due to the symmetry of the problem the solution of equation (\ref{AS}) must be 
symmetric on $y;{{\varphi}_1}'(y=0)=0$ due to $E_y=0$ at $y=0$.The solution 
of the linear problem \cite{H},\cite{I} is concentrated in the region of the 
"trace",falling outside it exponentially.
Adopting the same picture of the potential flow for considering case 
too,the solution of eq. (\ref{AS}) is 
$$\varphi_1(y)=\frac{ky^2}{2}+A=k\left(\frac{y^2}{2}+a\right),$$which is 
valid for small values of y.It means that,the solution of eq. (\ref{AP})
in separable arguments exists only for small values of $y<b$;

The equation (\ref{AT}) is the equation for 
nonlinear oscilator,with nonlinear part of the force proportional to 
$\varphi_2^2$ (for mathematical pendulum the first nonlinear term is 
proportional
to $\varphi_2^3$ see e.g. \cite{L}). The general solution of this 
equation is given in the implicit form by
\begin{equation}
\label{AW}
-(z+d)=\pm \int_{{\varphi}_2}^{{\varphi}_0}\frac{d\varphi_2}{\sqrt{2}{[h-
F(\varphi_2)]}^{1/2}}.
\end{equation}
sign $\pm$ corresponds to positive or negative $\frac{d\varphi_2}{dz}$ 
subsequently and $h$ is an energy constant,defined by 
\begin{equation}
\label{AX}
h=\frac{1}{2}{{\varphi'}_{2}^{2}}+\frac{n_0}{2}{\varphi}_{2}^{2}+\frac{k}{3}
{\varphi}_{2}^{3}
\end{equation}

and determined from boundary condition at $z=-d, {\varphi}_{2}(-d)\equiv
{\varphi}_0, {\varphi}_{2}'(-d)\equiv{\varphi}_{0}'$

The function $F({\varphi}_2)$ is
\begin{equation}
\label{AY}
F({\varphi}_2) =\frac{n_0}{2}{\varphi}_{2}^{2}+\frac{k}{3}
{\varphi}_{2}^{3}
\end{equation}
The separatrix,$h_s=\frac{1}{6k^2}n_0^3$,
is the tangent to $F(\varphi_2)$ at its maximum point;$F(\varphi_2)$
has three real roots:double root equal zero and one root at 
$B=\frac{3n_0}{2k}$
The roots of the equation
\begin{equation}
\label{AZ}
h-F({\varphi}_2)=0,
\end{equation}
are $c_i (i=1,2,3)$
The different solutions (\ref{AW}) of the equation (\ref{AT}) defined by 
the value of the $h$,which in turn,depends on the boundary values 
$\varphi_0$ and ${\varphi'}_0$.
Finite solutions  for $k<0$ is $c_1 \leq \varphi_2 \le c_2$,which is 
existed,when $h \le h_s$, i.e. $c_s \le c_1 \le 0$,where \\
$c_s=-\frac{n_0}{2|k|}$, and $0 \le c_2 \le c_m=
\frac{n_0}{|k|}$, \\
($c_m$ corresponds to the local maximum of the function 
$F(\varphi_2)$,equal to $h_s$).The third root of the eq (\ref{AZ})
is $c_3$: $c_m \le c_3 \le B = \frac{3n_0}{2|k|}$.For $k>0$ finite 
solutions exist,when $c_2 \le \varphi_2 \le c_3,0 \le h \le h_s$;
for $k=1, -\frac{3n_0}{2} \le c_1 \le -n_0,-n_0 \le c_2 \le 0,
0 \le c_3 \le \frac{n_0}{2};$

For the cylindrical bunch with length $2d$ and radius $R_0$ eq. (\ref{AJ})
is written in the form
$$\frac{\partial^2 \varphi}{\partial 
z^2}=\left[n_0+\frac{1}{r}\frac{\partial}{\partial r}\left(r\frac{\partial 
\varphi}{\partial r}\right)\right]V_z$$

The approximate equation of motion for $V_z$ has the same form as (\ref{AM})
and $V_z$ approximately is equal to $\varphi$.Basic equation for $\varphi$
is then $$\frac{\partial^2 \varphi}{\partial 
z^2}+\varphi\frac{1}{2}\frac{\partial}{\partial 
r}\left(r\frac{\partial \varphi}{\partial r}\right)+n_0\varphi=0$$.

Solution of this equation in separable arguments $\varphi(r,z)=\varphi_1(r)
\varphi_2(z)$ can be obtained by solving the equations:
$$\frac{1}{2}\frac{d}{dr}\left(r\frac{d\varphi_1}{dr}\right)=k$$
$$\frac{d^2\varphi_2}{dz^2}+n_0\varphi_2+k\varphi^2_2=0$$.

The last equations for $\varphi_2(z)$ coincides with the eq (\ref{AT})
for flat case.Equation for $\varphi_1(r)$ has the solution finite for small
$r$

$$\varphi_1(r)=\frac{k}{2}\left(\frac{r^2}{2}+a\right)$$.
Hence the cylindrical bunch case is described practically by the same 
equations as a flat one with the evident changes from $\varphi_1(y)$ to
$\varphi_1(r)$.

\section{\small{BOUNDARY CONDITIONS}}
\indent

The definition $\varphi=\varphi_1(y)\varphi_2(z)$ permits the transformation
$\varphi_1=|k|\bar{\varphi}_1,\varphi_2=|k|^{-1}\bar{\varphi}_2$.Then eqs.
(\ref{AS}) and (\ref{AT}) for $\bar{\varphi}_1$ and $\bar{\varphi}_2$ will 
have the following forms:
\begin{equation}
\label{EA}
\bar{\varphi}''_1=\pm 1,\bar{\varphi}''_2 +n_0\bar{\varphi}_2 \pm 
\bar{\varphi}_2^2=0,
\end{equation}
which corresponds to value of $k=\pm 1$.It means that separation constant $k$
is arbitrary and can be choisen as $k= \pm 1$.In what follows the sign 
"bar" over $\varphi_{1,2}$ is ommited and $k$ is choosen plus one,which 
provide meaningful results for our case,when $z<-d;k=-1$ is suitable for 
$z>0$.Then 
\begin{equation}
\label{EB}
\varphi=\varphi_1(y)\varphi_2(z)=(a+y^2/2)\varphi_2(z)
\end{equation}
If at $z=-d$ the physical quantities are
\begin{eqnarray}
\label{EC}
E_z^d\equiv E_z(y,z=-d)=E_{z0}+E_{z2}y^2=-{\left(\frac{\partial \varphi}
{\partial z}\right)}_{z=-d},\\ \nonumber
V_z^d \equiv V_z(y,z=-d)=V_{z0}+V_{z2}y^2=-\varphi(y,z=-d),
\end{eqnarray}
the unknown constants $\varphi_0,\varphi'_0,a$ entering in the solution
(\ref{AS},\ref{AW},\ref{AX},\ref{EB}) are
\begin{equation}
\label{ED}
\varphi'_0=-2E_{z2},\varphi_0=-2V_{z2},a=\frac{2E_{z0}}{E_{z2}}=
\frac{V_{z0}}{2V_{z2}}
\end{equation}
From (\ref{AG}.a) and (\ref{AT})
\begin{equation}
\label{EF}
NV_z=\frac{\partial^2 \varphi}{\partial z^2}=\varphi_1(y)
{\varphi''}_2(z)=\varphi_1(y)(-n_0\varphi_2-\varphi_2^2)
\end{equation}
and,when z=-d 
$${(NV_z)}^d=-(a+y^2/2)(\varphi_0+n_0)\varphi_0=V_z^d(\varphi_0+n_0),$$
i.e.
\begin{equation}
\label{EG}
N^d=\varphi_0+n_0,N^d=n_e^d(1+V_z^d)
\end{equation}
and is independent from $y$.

Consider the case of underdense plasma $n_b>n_0$,when all plasma electrons
behind the bunch are "blow out" $n_e^d=0$,i.e. $\varphi_0=-n_0$ 
\cite{M}.Following \cite{M} assume that
$E_z^d=0$ i.e. ${\varphi'}_0=0$ according to (\ref{EC},\ref{ED}).Then 
constant $h$ (\ref{AX}) is equal to 
$$h=\frac{n_0}{2}\varphi^2_0+\frac{1}{3}\varphi_0^3=\frac{1}{6}n_0^3=h_s$$
i.e. the solution,corresponding to the "blow out" regime, lies on separatrix.

The constant $h$ (\ref{AX}),also can be 
expressed through the roots $c_i=\alpha_i n_0$ of the equation (\ref{AZ}):

\begin{equation}
\label{BF}
h=\frac{1}{3}c_1c_2c_3=-\frac{{\alpha}_{1}{\alpha}_{2}{\alpha}_{3}}{3}
n_0^3
\end{equation}
where
$-3/2 \le \alpha_1 \le -1,-1 \le \alpha_2 \le 0,0 \le \alpha_3 \le 1/2$
for $k>0$.
For the separatrix $\alpha_1=\alpha_2=-1,\alpha_3=1/2$
\begin{equation}
\label{BG}
h=h_s=\frac{n_0^3}{6}
\end{equation}

For the values $h>h_s$ and $h<0$ as it is evident,the solutions for $\varphi_2$
have an infinite values.
When ${\varphi}_0<c_2$ the solution became unphysical,even for $0 \le h <h_s$

\section{\small{FINITE NONLINEAR SOLUTIONS}}
\indent

First consider the case when $c_2 \leq {\varphi}_0,{\varphi}_2<c_3$. From 
general
solution (\ref{AW}),using known expressions for the elliptic integrals and 
elliptic functions \cite{N},\cite{O},we have

\begin{equation}
\label{BK}
{\varphi}_2(z)=c_3-(c_3-c_2){sn}^{2}z_1,
\end{equation}

where

\begin{equation}
\label{BL}
z_1 \equiv F({\gamma}_0,q)+\frac{1}{2}\sqrt{\frac{2(c_3-c_1)}{3}}(z+d)
\end{equation}

\begin{equation}
\label{BM}
{\gamma}_0=\arcsin{\sqrt{\frac{c_3-\varphi_0}{c_3-c_2}}},
\end{equation}
\begin{equation}
\label{K}
q=\sqrt{\frac{c_3-c_2}{c_3-c_1}}=\sqrt{\frac{{\alpha}_3-{\alpha}_2}
{{\alpha}_3-{\alpha}_1}}
\end{equation}

and $F({\gamma_0},q)$ is the elliptic integral of the first kind,$snz_1$-
elliptic function.

Using (\ref{AJ}),(\ref{EB}),(\ref{ED}) and 
(\ref{BK})-(\ref{BM}) it is possible to obtain
\begin{equation}
\label{EH}
E_y=-\frac{\partial \varphi}{\partial y}=-y\varphi_2(z)=
-yn_0[\alpha_3-(\alpha_3-\alpha_1)sn^2z_1]
\end{equation}
\begin{eqnarray}
\label{EI}
E_z=-\frac{\partial \varphi}{\partial z}=\left(\frac{y^2}{2}+a\right)n_0
(\alpha_3-\alpha_2)\left[\frac{2}{3}n_0(\alpha_3-\alpha_1)\right]^{1/2} 
sn z_1ch z_1dn z_1= \\ \nonumber
=\frac{1}{2}\left(\frac{y^2}{2}+a\right)n_0
(\alpha_3-\alpha_2)\left[\frac{2}{3}n_0(\alpha_3-\alpha_1)\right]^{1/2}
sn2z_1\left[1-\left(\frac{\alpha_2-\alpha_1}{\alpha_3-\alpha_1}\right)
sn^4z_1\right]
\end{eqnarray}

The length $\lambda_n$ of the nonlinear wave is given by 
\begin{eqnarray}
\label{BS}
\frac{\omega}{v_0}{\lambda}_n=2\int_{c_2}^{c_3}\frac{d\varphi_2}
{\sqrt{2}{[h-F({\varphi}_2)]}^{1/2}}; \\ \nonumber
{\lambda}_n=\frac{4v_0}{\omega}{\left(\frac{3}{2n_0}\right)}^{1/2}
\frac{1}{{({\alpha}_3+|{\alpha}_1|)}^{1/2}} \times \\ \nonumber
\times F\left(\frac{\pi}{2},{\left(
\frac{{\alpha}_3+|{\alpha}_2|}{{\alpha}_3+|{\alpha}_1|}\right)}^{1/2}\right)
\end{eqnarray}
In the linear case 
$$h \rightarrow 0,\alpha_2
\rightarrow 0 \alpha_3 \rightarrow 0,\alpha_1 \rightarrow - 3/2$$ and
$\lambda_n \rightarrow \lambda_p=\frac{2{\pi}v_0}{\omega_p}$;

From (\ref{EH}) it follows,that $E_y$ is by the order of magnitude equal 
to $E_y \sim yn_0$,or in the ordinary units $$E_y \sim \frac{m\omega_pv}{e}
\frac{\omega_p}{v_0}y=4\pi en_py,$$ which coincides with the field inside
the flat bunch uniformly charged with the plasma electron density $n_p$.

From (\ref{EI}) the longitudinal component of the electric field is by 
order of magnitude equal to $E_z \sim \left(\frac{y^2}{2}+a\right)n_0^{3/2}$,
or in ordinary units $$E_z \sim 
\frac{m\omega_pv_0}{e}\frac{\omega_p^2}{v_0^2}\left(\frac{y^2}{2}+\tilde{a}
\right)=4\pi en_p\frac{\omega_p}{v_0}\left(\frac{y^2}{2}+\tilde{a}\right)$$
where $\tilde{a}=\frac{2E_{z0}}{E_{z2}}, E_z(y,z=-d)=E_{z0}+E_{z2}y^2$ in
ordinary units,so at some conditions it can be larger than $E_y$.

In the case $h=h_s$ the changes due to nonlinarity are more drastic.

As we have seen,this case corresponds to the conditions $$n_b >n_0,n_e(z=-d,
y=0)\equiv n_d=0,$$ and $E_{z0}=0$, which resembles the blow-out regime in
underdense plasma (\ref{AH}).
In this case $$c_1=c_2\rightarrow c_m=-n_0,c_3\rightarrow c_s=+n_0,$$
$$h-F(\varphi_2)\rightarrow\frac{1}{3}(\varphi_2+n_0)^2\left(\frac{n_0}{2}-
\varphi_2\right)$$

From (\ref{AW}) with the minus sign in the front of integral ($\varphi_2$
decreases,when $z$ increases from $-\infty$ up to $-d$),using \cite{N} it 
follows 
\begin{equation}
\label{EJ}
\varphi_2=-n_0+3\frac{n_0}{2}th^2\frac{\psi}{2}
\end{equation}
where
\begin{equation}
\label{EK}
\psi=\theta_0-\sqrt{n_0}(z+d), \theta_0=ln\left|\frac{\left(\frac{n_0}{2}-
\varphi_2\right)^{1/2}+\left(\frac{3}{2}n_0\right)^2}{\left(\frac{n_0}{2}-
\varphi_0\right)^{1/2}-\left(\frac{3}{2}n_0\right)^2}\right|
\end{equation} 
In the (\ref{EK}) it is necessary to use $\varphi_0=-n_0+\epsilon,
\epsilon > 0$and pass to limit $\epsilon \rightarrow 0$ at fixed $\psi$.
Then at $\psi = 0,z \rightarrow -\infty$,and at $\psi \rightarrow -\infty,
z\rightarrow -d$.

The electric field components in the considered case of the blow-out 
regime are 
\begin{equation}
\label{EL}
E_y=\frac{1}{2}yn_0(3th^2\psi/2-1)
\end{equation}
\begin{equation}
\label{EM}
E_z=3n_0^{3/2}\left(\frac{y^2}{2}+\frac{V_{z0}}{n_0}\right)(1-th^2\psi/2)
th\psi/2
\end{equation}

The maximum value of the transversal component of the field by the order of
magnitude is $E_y^{max}\sim yn_0$.

Longitudinal component $E_z=0$,when $z\rightarrow -d , (\psi \rightarrow
-\infty)$ and when $z\rightarrow -\infty , (\psi \rightarrow 0)$.
Its maximum value is at $\psi_0=-1,32$ and by the order of magnitude is equal
$$E_z^{max} \sim n_0^{3/2}\left(\frac{y^2}{2}+\frac{V_{z0}}{n_0}\right).$$
Hence the maximum values of the field coincides with that in case when
$0 \leq h < h_s$.The difference is in the form of the wave:when $h < h_s$
the wake wave  (\ref{EH}-\ref{EI}) is cnoidal and when $h=h_s$ (blow out 
regime) the wave (\ref{EL}-\ref{EM}) is solitary one.

\section{\small{APPROXIMATE SOLUTION OBTAINED BY MULTIPLE SCALES METHOD}}

In order to obtain the solution of the basic eq. (\ref{AP}) valid for larger
values of $y$ it is necessary to solve (\ref{AP}) by some other 
methods.It seems that the multiple scales approximate method (development 
of the derivative \cite{P}) is suitable for this propose.

The small parameter in question for considered case is $\epsilon 
=\left(\frac{e\varphi_{max}}{mc^2}\right) \ll 1 , (\varphi' \ll 
1,\varphi$ here is in ordinary units).According to \cite{P} introduce 
the different scales variables in $z$
$$ z_0=z,z_n=\epsilon^n(z+d),n=1,2,3 \dots$$
and perform the following developments 
\begin{eqnarray}
\label{EN}
\varphi(y,z)=\tilde{\varphi}(y,z_0,z_1,z_2,\dots)= && \\ \nonumber
=\epsilon\varphi_1
(y,z,z_1,z_2,\dots)+\epsilon^2\varphi_2(y,z,z_1,z_2,\dots)
+\epsilon^3\varphi_3(y,z,z_1,z_2,\dots) \\ \nonumber
\frac{\partial}{\partial z}=\frac{\partial}{\partial z_0}+\epsilon
\frac{\partial}{\partial z_1}+\epsilon^2\frac{\partial}{\partial z_2}+\dots 
&& \\ \nonumber
\frac{\partial^2}{\partial z^2}=\frac{\partial^2}{\partial z_0^2}+
2\epsilon\frac{\partial^2}{\partial z_0\partial z_1}
\epsilon^2\left(2\frac{\partial^2}{\partial z_0\partial 
z_1}+\frac{\partial^2} {\partial z_1^2}\right)+\dots
\end{eqnarray}
(For dimensionless function and arguments it is necessary to put 
$\epsilon =1$ in the final results.)

Substitution of the developments (\ref{EN}) in (\ref{AP}) gives the following
set of the equations for subsequent approximations:

\begin{equation}
\label{EO}
\frac{\partial^2\varphi_1}{\partial z_0^2}+n_0\varphi_1=0
\end{equation}
\begin{equation}
\label{EP}
\frac{\partial^2\varphi_2}{\partial z_0^2}+n_0\varphi_2=
-\varphi_1\frac{\partial^2\varphi_1}{\partial y^2}-
2\frac{\partial^2\varphi_1}{\partial z_0\partial z_1}
\end{equation}
\begin{equation}
\label{ER}
\frac{\partial^2\varphi_3}{\partial z_0^2}+n_0\varphi_3=
-\varphi_1\frac{\partial^2\varphi_2}{\partial y^2}-\varphi_2
\frac{\partial^2\varphi_1}{\partial y^2}-2\frac{\partial^2\varphi_2}
{\partial z_0 \partial z_1}-\left(2\frac{\partial^2}
{\partial z_0 \partial z_2}+\frac{\partial^2}{\partial z_1^2}\right)\varphi_1
\end{equation}

The general solution of eq. (\ref{EO}) is 
\begin{equation}
\label{ES}
\varphi_1=a(y_1,z_1,z_2)e^{-i\sqrt{n_0}z_0}+a^*(y_1,z_1,z_2)
e^{+i\sqrt{n_0}z_0}
\end{equation}
and $y$-dependence of the solution (\ref{ES}) comes from boundary conditions
at $z_0=-d_1,z_1=0,z_2=0$.If $$E_z=-\frac{\partial 
\varphi_1(y,z_0=-d,z_1=0,z_2=0)}{\partial z_0} \equiv g(y)$$
$$E_y=-\frac{\partial \varphi_1(z_0=-d,z_1=0,z_2=0)}{\partial y}\equiv f(y)$$
then 
\begin{eqnarray}
\label{ET}
Re a(y,z_0=-d,z_1=0,z_2=0)=\\ \nonumber
=-\frac{1}{2}\int_0^y \left(f(y)\cos n_0d+\frac{1}{\sqrt{n_0}}g(y)
\sin\sqrt{n_0}d\right)dy \\ \nonumber
Im a(y,z_0=-d,z_1=0,z_2=0)=\\ \nonumber
=\frac{1}{2}\int_0^y \left(f(y)\sin \sqrt{n_0}d+\frac{1}{\sqrt{n_0}}g(y)
\cos\sqrt{n_0}d\right)dy
\end{eqnarray}
The second term is right hand side of the eq. (\ref{EP}) is secular due to
the solution (\ref{ES});it can be eliminated if $\varphi_1$ is independent
on $z_1$ i.e. 
\begin{equation}
\label{EU}
\frac{\partial \varphi_1}{\partial z_1}=0
\end{equation}

It means that $\varphi_2$ is also independent on $z_1$ and the general 
solution of the eq. (\ref{EP}),taking into account (\ref{EU}),is 
\begin{equation}
\label{EV}
\varphi_2=B(y,z_2)+A(y,z_2)e^{-2i\sqrt{n_0}z_0}+A^*(y,z_2)e^{2i\sqrt{n_0}z_0}+
b(y,z_2)e^{-i\sqrt{n_0}z_0}+b^*(y,z_2)e^{i\sqrt{n_0}z_0}
\end{equation}
where
\begin{equation}
\label{EW}
B=-\frac{1}{n_0}\left(a\frac{\partial^2a^*}{\partial y^2}+
a^*\frac{\partial^2 a}{\partial y^2}\right), A=-\frac{1}{3n_0}a
\frac{\partial^2 a}{\partial y^2}
\end{equation}

Function $b(y,z_2)$ entering in the solution of the homogenious part of the
eq.(\ref{EP}) can be found from the boundary conditions
$$\varphi_2(y,z_0=-d,z_2=0)=0$$
$$\frac{\partial \varphi_2(y,z_0=-d,z_2=0)}{\partial z_0}=0$$ and has the
following value
\begin{equation}
\label{EX}
b=\frac{1}{2}\left(A^*e^{-3i\sqrt{n_0}d}-3Ae^{i\sqrt{n_0}d}+B\right)
\end{equation}

The $z_2$-dependense of the function $a(y,z_2)$ (and subsequently the 
$z_2$-dependense of $A,B,b$) comes out from the consideration of the eq. 
(\ref{ER}) for the third approximation.Eq. (\ref{ER}),due to the 
independense of $\varphi_1,\varphi_2$ on $z_2$ (\ref{EU}),has the form:
\begin{equation}
\label{EY}
\frac{\partial^2\varphi_3}{\partial z_0^2}+n_0\varphi_3=-2\frac{\partial^2
\varphi_1}{\partial z_0\partial z_2}-\varphi_1\frac{\partial^2
\varphi_2}{\partial y^2}-\varphi_2\frac{\partial^2\varphi_1}{\partial y^2}
\end{equation}

Using solutions (\ref{ES},\ref{EV}) for $\varphi_1$ and $\varphi_2$ it is
evident that right hand side of the eq. (\ref{EY}) has the secular terms,
proportional to $e^{\pm i\sqrt{n_0}z_0}$.The conditions for their 
elimination are 
\begin{eqnarray}
\label{EZ}
\frac{\partial a}{\partial z_2}-\frac{1}{n_0}
\frac{\partial^2 a}{\partial y^2}\left(a^*\frac{\partial^2 a}{\partial y^2}+
a\frac{\partial^2 a^*}{\partial y^2}\right)+\frac{1}{3n_0}a
\frac{\partial^2 a}{\partial y^2}\frac{\partial^2 a^*}{\partial y^2}- \\ 
\nonumber
-\frac{1}{n_0}a\frac{\partial^2}{\partial y^2}\left(a^*\frac{\partial^2 a}
{\partial y^2}+a\frac{\partial^2 a^*}{\partial y^2}\right)+\frac{1}{3n_0}
a^*\frac{\partial^2}{\partial y^2}\left(a\frac{\partial^2 a}{\partial y^2}
\right)
\end{eqnarray}
and subsequent conjugate expression.

The eq. (\ref{EZ}) determines the dependense of $a(y,z_2)$ from $z_2$.Eq. 
(\ref{EZ}) is complicated enough,it is a system of the first order 
differential equations for $Re a(y,z_2)$ and $Im a(y,z_2)$. The 
$y$-dependense of $a(y,z_2=0)$ is given by boundary conditions.

The eq. (\ref{EZ}) simplifies under the assumption that $a(y,z_2)$ is
\begin{equation}
\label{A}
a(y,z_2)=Y(y)Z(z_2)
\end{equation}
where $Z(z_2=0)=1$,and $Y(y)$ is known from the boundary conditions 
(\ref{ET}).Under (\ref{A}) eq. (\ref{EZ}) takes the form
\begin{equation}
\label{B}
2i\sqrt{n_0}\frac{dz}{dz_2}=\psi(y)|Z|^2Z
\end{equation}
where $\psi(y)$ is
\begin{eqnarray}
\label{C}
\psi(y) \equiv \frac{8}{3n_0}Y''{Y^*}''*+\frac{2}{3n_0}\frac{Y^*}{Y}
({Y''}^2-Y'Y'''-\frac{1}{2}YY'''') \\ \nonumber
+\frac{2}{n_0}({Y^*}'Y'''+Y'{Y^*}'''+\frac{1}{2}Y^*Y''''+\frac{1}{2}Y{Y^*}'''')
\end{eqnarray}

The function $\psi(y)$ is slowly varied on $y$,when $|y| \leq b$ and is 
zero,when $|y|>b$.It is reasonable to average the eq. (\ref{B}) on $y$.
Strikly speaking the need of such kind of procedure indicates that assumption
(\ref{A}) is not in full appropriate,but practically can work for slowly
varying $\psi(y)$.

After averaging,eq. (\ref{B}) and its conjugate one give the following 
system for $x \equiv ReZ$ and $y \equiv ImZ$ 
\begin{equation}
\label{D}
2\sqrt{n_0}\frac{dx}{dz_2}=Im\bar{\psi}|Z|^2x+Re\bar{\psi}|Z|^2y
\end{equation}
\begin{equation}
\label{E}
-2\sqrt{n_0}\frac{dy}{dz_2}=Re\bar{\psi}|Z|^2x-Im\bar{\psi}|Z|^2y
\end{equation}
where
$$\bar{\psi} \equiv \frac{1}{b}\int_0^b\psi(y)dy.$$ From (\ref{D}-\ref{E})
follows
\begin{equation}
\label{F}
\frac{d|Z|^2}{dz_2}=\frac{2Im\bar{\psi}}{\sqrt{n_0}}|Z|^4.
\end{equation}
The eq. (\ref{F}) has the solution
\begin{equation}
\label{G}
|Z|^2=\frac{1}{c-\frac{2Im\bar{\psi}z_2}{\sqrt{n_0}}} \geq 0,
\end{equation}
with the arbitrary constant $c>\frac{2Im\bar{\psi}}{\sqrt{n_0}}$.In the 
simplest case,when $Im\bar{\psi}=0 , |Z|^2=c^{-1}$ and the system (\ref{D}-
\ref{E}) has the solution
\begin{eqnarray}
\label{H}
x=c^{-1/2}\cos\left[\frac{Re\bar{\psi}}{2\sqrt{n_0}c}z_2+\theta_0\right] \\
\nonumber
y=-c^{-1/2}\sin\left[\frac{Re\bar{\psi}}{2\sqrt{n_0}c}z_2+\theta_0\right]
\end{eqnarray}

The constants $c^{-1/2}=1,\theta_0=0$,due to the boundary condition 
$Z(z_2=0)=1$.When $Im\bar{\psi} \neq 0,x=|Z|\cos\theta,y=-|Z|\sin\theta$,
where $\theta$ is the solution of equation
\begin{equation}
\label{I}
\frac{d\theta}{dz_2}=-\frac{|Z|^2}{2\sqrt{n_0}}Re\bar{\psi}
\end{equation}
at boundary condition $Z(z_2=0)=1,(\theta=0)$.
\begin{equation}
\label{J}
\theta=\frac{1}{4}\frac{Re\bar{\psi}}{Im\bar{\psi}}\ln(1-\frac{2Im\bar{\psi}}
{\sqrt{n_0}}z_2)
\end{equation}
and $a(y,z_2)$ for the first approximation (\ref{ES}),is $a(y,z_2)=
Y(y)Z(z_2)=Y(y)|Z|e^{i\theta}$,where $|Z|$ is given by (\ref{G}),$\theta$
by (\ref{J}) and $Y(y)$ can be found from boundary condition (\ref{ET}).
When $Im\bar{\psi}\rightarrow 0$ solution (\ref{G}),(\ref{J}) turns to
the solution (\ref{H}).

The solution (\ref{J}) is valid for $|z_2|=\epsilon^2|z'|=\left(\frac
{e\varphi_{max}}{mc^2}\right)^2|z'| \leq 1$ i.e. for $|z'|=k_p|z| 
\leq\left(\frac
{mc^2}{e\varphi_{max}}\right)^2$.In the  considered domain  $z_2 < 0,(z < 
-d)$ and the solutions (\ref{G}),(\ref{J}) have  different behaviour,when 
$Im\bar{\psi} > 0$ and $Im\bar{\psi} < 0$.When $Im\bar{\psi} > 0$ from 
(\ref{G})
it is seen that $|Z|^2$ (and consequently the amplitude $a(y,z_2)$) 
decreases when $|z_2|$ increases;when $Im\bar{\psi}<0,|Z|^2$ and 
amplitude $a(y,z_2)$ increases,when $|z_2|=\epsilon^2|z|$ increases 
up to allowed value:
$$|z|_{max}=min\left\{\left(\frac{mc^2}{e\varphi_{max}}\right)^2,\left(\frac{mc^2}
{e\varphi_{max}}\right)^2\frac{\sqrt{n_0}}{2Im\bar{\psi}}\right\}$$.

An outlined approximate procedure,based on multiple scales method for solving
basic nonlinear equation (\ref{AP}) for potential wake waves,shows that in
the lowest (first,second and third) approximations solution of eq. (\ref{AP})
is represented by set of harmonics $\sim e^{i\sqrt{n_0}mz} (m=0,1,2,3 \dots)$
with amplitudes ($a,B,A,b$ eqs.(\ref{EV}-\ref{EX}), 
(\ref{A}),(\ref{G}-\ref{J})
slowly increasing or decreasing with $|z|$.The $z$-dependense 
of amplitudes is rather combersome.but it is clear,that some singularity can
appear at certain boundary conditions when $Im\bar{\psi}<0$ and $|z_2|\sim
\frac{\sqrt{n_0}}{2Im\bar{\psi}} \leq 
\left(\frac{mc^2}{e\varphi_max}\right)^2$,which is inside of the region 
of applicability of the adopted procedure.

May be it is an indication of some new quality of the considered nonlinear
potential wake wave and, if it is so, it needs an additional consideration.
In any case the oulined consideration indicates that boundary conditions 
at the rear end of the driving bunch,which can be changed by appropriate 
choise
of the bunch transverse and longitudinal charge distributions,can essentially
effect the kind of the nonlinear wave amplitude dependense on the 
longitudinal coordinate $z$,even in the first approximation.

It seems,that subsequent experimental investigation of the dependense of 
the wake wave amplitude on the driving bunch transverse and longitudinal 
charge distributions could  be useful.

\section{\small{ACKNOWLEDGEMENTS}}
\indent
 
Author would like to thank A.M. Sessler for attention and essential support,
S.S Elbakian for the valuable comments,A.G. Khachatryan for usefull discussion,
Cathy Vanecek for attention and care and Gayane Amatuni for the help in 
preparing the manuscript for publication.The work was supported by the 
International Science and Technology Center and Minatom of RF.

\end{document}